\documentclass[conference]{IEEEtran}
\IEEEoverridecommandlockouts
\usepackage{cite}
\usepackage{amsmath,amssymb,amsfonts}
\usepackage{algorithmic}
\usepackage{graphicx}
\usepackage{textcomp}
\usepackage{xcolor}
\usepackage{multirow}
\def\BibTeX{{\rm B\kern-.05em{\sc i\kern-.025em b}\kern-.08em
    T\kern-.1667em\lower.7ex\hbox{E}\kern-.125emX}}
\begin{document}

\title{\huge Extending 2D Saliency Models for Head Movement Prediction in 360-degree Images using CNN-based Fusion}
\author{\IEEEauthorblockN{Ibrahim	Djemai$^{\S \star}$ \ Sid Ahmed Fezza$^{\S}$ \ Wassim Hamidouche$^{\star}$ \ and \ Olivier D\'eforges$^{\star}$}
			    \textit{$^{\S}$National Institute of Telecommunications and ICT, Oran, Algeria}\\
					\textit{$^{\star}$Univ. Rennes, INSA Rennes, CNRS, IETR - UMR 6164, Rennes, France}}

\maketitle

\begin{abstract}
Saliency prediction can be of great benefit for 360-degree image/video applications, including compression, streaming, rendering and viewpoint guidance. It is therefore quite natural to adapt the 2D saliency prediction methods for 360-degree images. To achieve this, it is necessary to project the 360-degree image to 2D plane. However, the existing projection techniques introduce different distortions, which provides poor results and makes inefficient the direct application of 2D saliency prediction models to 360-degree content. Consequently, in this paper, we propose a new framework for effectively applying any 2D saliency prediction method to 360-degree images. The proposed framework particularly includes a novel convolutional neural network based fusion approach that provides more accurate saliency prediction while avoiding the introduction of distortions. The proposed framework has been evaluated with five 2D saliency prediction methods, and the experimental results showed the superiority of our approach compared to the use of weighted sum or pixel-wise maximum fusion methods.
\end{abstract}

\section{Introduction}
\label{sec:intro}
Omnidirectional or 360-degree image/video represents an essential type of virtual reality (VR) content which meets an increasing success. This type of content offers an immersive experience to viewers with the use of head mounted displays (HMDs), providing a limited field of view and three degrees of freedom (roll, yaw and pitch). This allows the viewer to explore the scene by freely turning his head, as humans do in the real world. Being close to the viewer's eyes, the HMD must provide high-resolution content (4K or above) in order to offer high quality of experience. Furthermore, to avoid motion sickness, the frame rate should also be high \cite{chen2018recent}. Therefore, in sight of the huge amount of data created by such omnidirectional images (ODIs), their storage and transmission over existing bandwidth-limited infrastructures pose a great challenge.

Since only a fraction of 360-degree images can catch the viewer attention, namely the \textit{viewport}, prediction of head movement in ODI is very useful to overcome the aforementioned limitations, and can be exploited in different applications, including compression, streaming, rendering and viewpoint guidance. For instance, compressing the non-viewed parts of the scene using high compression level is a way of reducing the amount of data sent over the network, and it is done by predicting where the user is most likely going to rotate his head, \textit{i.e.}, prediction of head fixations. 

\begin{figure}[t!]
    \centering
    \includegraphics[scale=0.18]{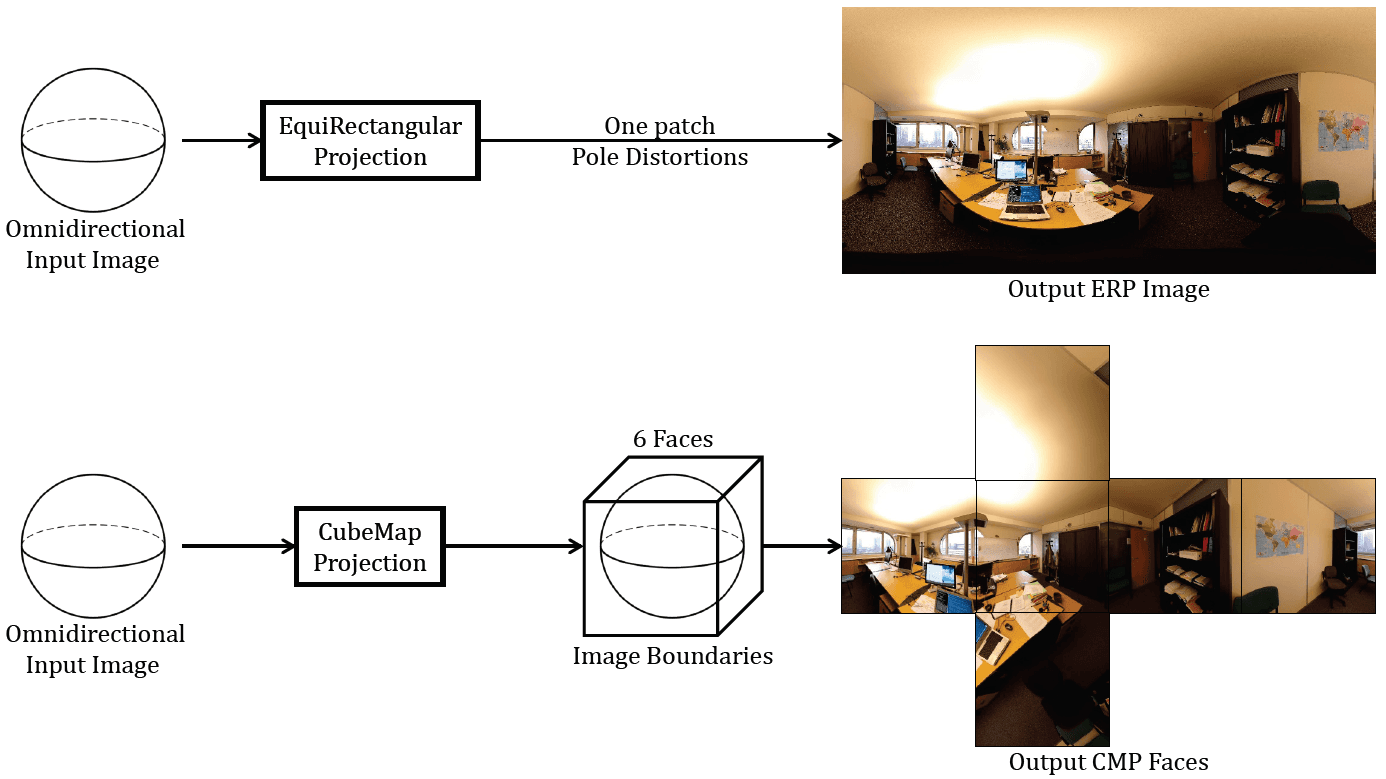}
    \caption{Examples of equirectangular (top) and cubemap (bottom) projections for omnidirectional content.}
    \label{figs1}
\end{figure}

Some works have proposed saliency models suited for this kind of content. These methods can be grouped under two different approaches: those who perform the prediction of saliency directly on the sphere \cite{zhang2018saliency}, and those attempting to extend the existing 2D saliency prediction methods to 360-degree content by applying them to one or more omnidirectional content projections \cite{chao2018salgan360, de2017look, lebreton2018gbvs360, maugey2017saliency, startsev2018360, battisti2018feature, monroy2018salnet360,zhu2018prediction,fang2018novel,biswas2017towards}, \textit{e.g.}, EquiRectangular Projection (ERP) or CubeMap Projection (CMP). 

The ODI is shown to the user through the HMD as a sphere, but it has to be stored as a regular 2D content in order to be processed. Many projections exist to transform the ODI from spherical format to a flat one \cite{li2019state}, and the ERP and CMP projections are the widely used ones for saliency prediction on ODIs, as illustrated in Figure \ref{figs1}. The ERP is represented with a single patch, but it suffers from geometric distortions in the poles, making it inappropriate to detect salient regions directly, since it is different from what observers really perceive. While CMP content suffers less from this issue, but it introduces image boundaries being represented with 6 faces, \textit{i.e.}, border artifacts. In other words, given the discontinuities between faces, applying the 2D saliency prediction method independently on the 6 faces takes into account only the local context of the scene, thus providing insufficient results.  

Thus, in order to compensate these geometric distortion and border artifacts due to sphere-to-plane projections, different approaches have been proposed. Some works proposed to combine both ERP and CMP projections \cite{startsev2018360,chao2018salgan360}. For instance, authors in \cite{chao2018salgan360} used the CMP projection to estimate local saliency maps, while the ERP projection has been exploited for global context saliency prediction. Other approaches proposed to apply a single projection, ERP or CMP, but with different rotations \cite{de2017look,lebreton2018gbvs360,maugey2017saliency,startsev2018360}. This rotation procedure is used in order to take into account only the distortion-free regions of the ERP, \textit{i.e.}, middle portion, or to deal with the border artifacts in the CMP projection.

However, all these approaches generate a set of intermediate saliency maps, resulting from different rotations or from different projection procedures \cite{chao2018salgan360, de2017look, lebreton2018gbvs360, maugey2017saliency, startsev2018360, battisti2018feature, monroy2018salnet360,zhu2018prediction,fang2018novel}, which requires a fusion block to merge them to provide a single final saliency map. This is usually done either by a weighted sum or a pixel-wise maximum between the resulted saliency maps. In addition, this fusion is performed after a re-projection of the saliency maps to ERP format. Thereby, these approaches are static and do not provide an optimal result.     
 


Given the rich literature and the significant advances reached by the 2D saliency prediction approaches, consequently, in this paper, we propose a novel framework to  effectively extend any 2D saliency prediction method for 360-degree images. To reach this goal, as a first step, the 2D saliency prediction method is applied on cube face images (CMP images for short) under several orientations. Then, in contrast to conventional approaches, the resulting saliency maps are optimally merged using a convolutional neural network (CNN) based fusion approach. This CNN has been trained with CMP images using a new loss function including different saliency evaluation metrics. This framework is modular and can be used with any 2D saliency prediction approach. The experiments with five 2D saliency prediction methods have shown that the proposed adaptation method outperforms baseline methods using a weighted sum or pixel-wise maximum.
 \begin{figure*}[t!]
		\centering
		\includegraphics[scale=0.31]{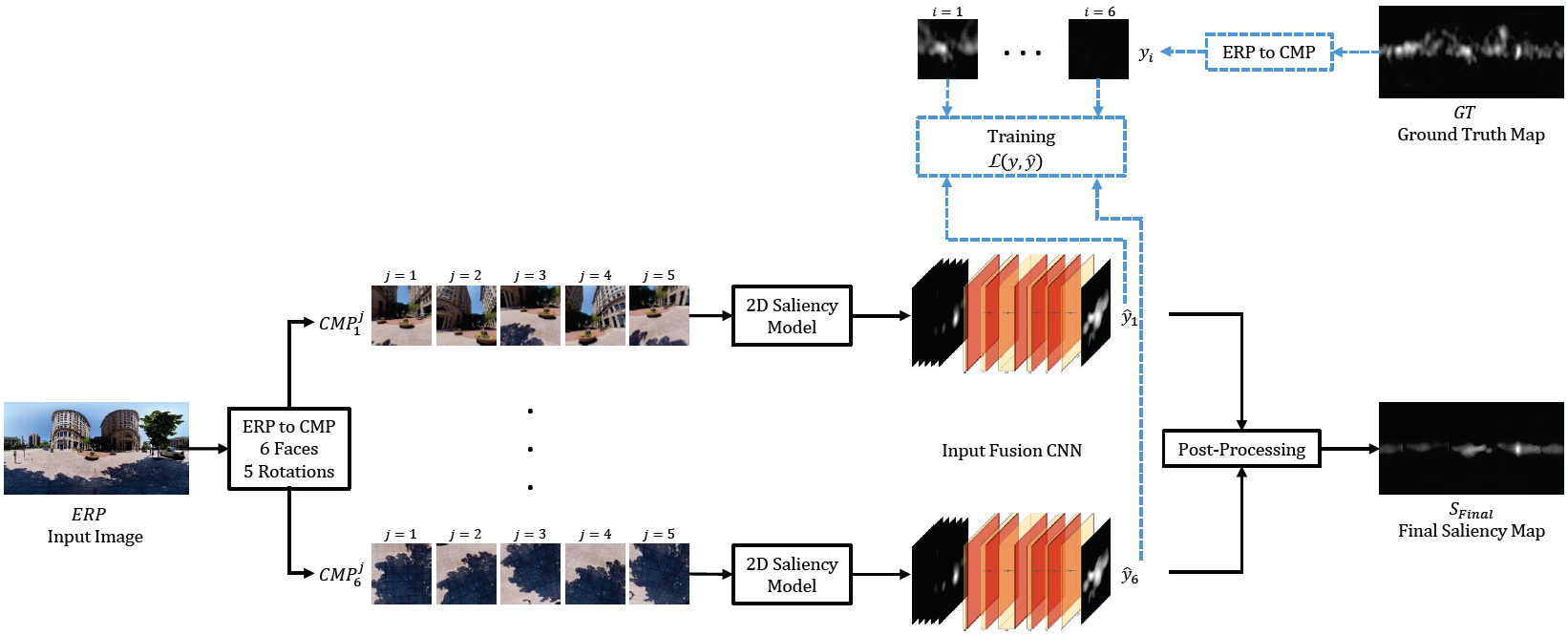} 
		\caption{Architecture of the proposed framework. The blue dotted lines represent the training procedure.}
		\label{fig:framework}
	\end{figure*}

The rest of the paper is organized as follows. Section \ref{sec:related} presents a review of the saliency prediction methods for 360-degree images. Section \ref{sec:proposal} describes the proposed approach. The experimental results are presented in Section \ref{sec:experimental}. Finally, Section \ref{sec:conclusion} concludes the paper.
\section{Related Work}
\label{sec:related}
Most of the methods proposed to predict the visual saliency of 360-degree content are extensions of 2D saliency models. The difference between them lies mainly in the choice of the type of sphere-to-plane projection and, if the exploited 2D saliency method is hand-crafted or deep neural features based approach.

For instance, Abreu \textit{et al.} \cite{de2017look} proposed to apply a 2D saliency prediction method on several ERP images obtained with four horizontal translations. Then, the obtained saliency maps are fused using a weighted averaging to derive the final saliency map of the 360-degree image. Inspired by this work, Lebreton \textit{et al.} \cite{lebreton2018gbvs360} proposed an extension of two 2D saliency models,  Graph-Based Visual Saliency (GBVS) \cite{harel2007graph} and Boolean Map Saliency (BMS) \cite{zhang2013saliency}, called GBVS360 and BMS360, in which the same fusion procedure proposed in \cite{de2017look} is used.

In order to avoid the border artifacts due to the CMP format, in \cite{maugey2017saliency}, the authors proposed to use a double cube projection, which consists of projecting the spherical image into two cubes, such that the cube face centers are pointing towards the corners of the other cube. Next, the 2D saliency is estimated for all faces of both cubes and are projected back to the ERP plane and then fused using a weighted average. The 2D saliency map of each face of each cube is obtained by aggregating the saliency maps computed by 5 state-of-art models, including GBVS \cite{harel2007graph}, Image Signature (ImgSig) \cite{hou2011image}, Adaptive Whitening Saliency (AWS) \cite{garcia2012saliency} and BMS \cite{zhang2013saliency}.

In contrast to the previous described works that exploited only one type of projection, Startsev \textit{et al.} \cite{startsev2018360} considered both CMP and ERP projections for the saliency prediction of 360-degree images. Specifically, they used two ERP images in addition to two CMP faces generated at several different rotations. Then, the saliency maps of CMP images are projected back to the ERP format. Finally, all ERP saliency maps of all orientations are fused together through pixel-wise maximum. In the same vein, Chao \textit{et al.} \cite{chao2018salgan360} proposed to use the original ERP image to derive the global saliency map, while CMP images with different rotations have been used to generate the local saliency maps. Finally, the global and local saliency maps are linearly combined to generate the final saliency map in the ERP format.

In \cite{battisti2018feature}, based on the viewport images, both low- and high-level feature maps are generated using low-level features and high-level or semantic features. Finally, both maps are fused in order to obtain the final saliency map. Similarly, in \cite{zhu2018prediction}, the authors proposed a combination between the output of bottom-up and top-down saliency maps, which operate on a special projection that simulates the user's viewport. Fang \textit{et al.} \cite{fang2018novel} proposed to predict the saliency map of 360-degree image by combining the feature contrast and boundary connectivity. The feature contrast is computed on superpixel level by low-level features, while the boundary connectivity is defined as background measure to extract background map.
\section{Proposed Approach}
\label{sec:proposal}
This section presents the proposed framework, as shown in Figure \ref{fig:framework}, which includes cubemap projections, application of the 2D saliency prediction model, CNN-based fusion and post-processing. The details of each of these steps are provided below.
\subsection{Cubemap Projections}
\begin{figure}[t!]
    \centering
    \includegraphics[scale=0.2]{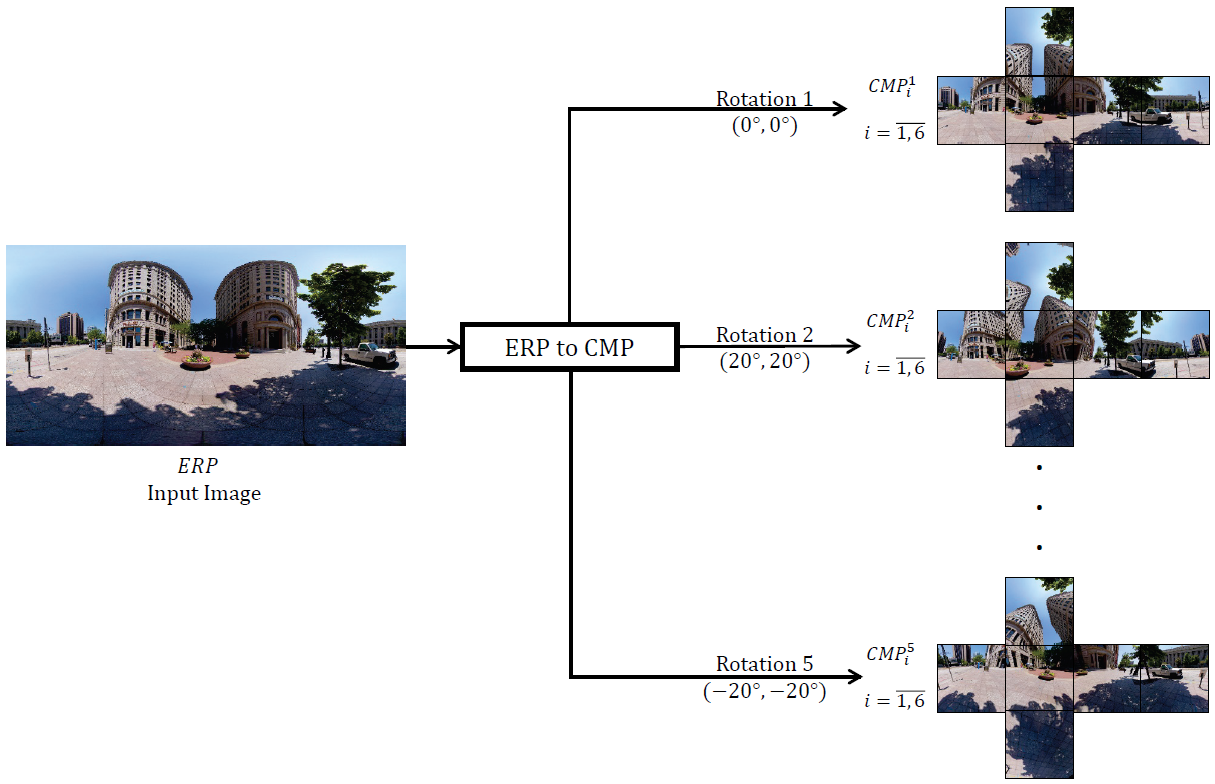}
    \caption{Cubemap projections with 5 different rotations.}
    \label{CubemapProj}
\end{figure}

Since the ERP suffers from geometric distortions, applying a regular 2D saliency model on it would produce poor results. Therefore, the input ERP image is converted into CMP faces. However, CMP introduces image boundaries to the resulting cube faces. To deal with this issue, we perform several CMPs with different rotations. As shown in Figures \ref{fig:framework} and \ref{CubemapProj}, we have chosen 5 different rotations (in yaw and pitch) of the ERP input image to be converted to CMP, which are respectively \hbox{(0$^\circ$, 0$^\circ$),} (20$^\circ$, 20$^\circ$), (-20$^\circ$, 20$^\circ$), (20$^\circ$, -20$^\circ$) and (-20$^\circ$, 20$^\circ$). This setup insures covering all directions with respect to the central rotation (0$^\circ$, 0$^\circ$).

More formally, given an input equirectangular image $ERP \in \mathbb{R}^{c\times w\times h}$, where $c$ is the number of channels, while $w$ and $h$ represent the width and height, respectively. The projection function $P$ transforms the $ERP$ to a cubemap format $CMP^{j}\in \mathbb{R}^{6\times c \times l\times l}$ under the rotation $j$, where $l$ is the edge length of the cube and $CMP^{j}$ represents a stack of six faces. Given the projections have been performed using 5 different rotations, this provides a set of CMP images, where \hbox{$CMP=\{ CMP^{j}_i | i=1,...,6, j=1,...,5\}$.}  

\subsection{2D saliency model}
Once the CMP faces have been generated, we apply on them a regular 2D saliency model. We have considered 5 models to evaluate our appraoch, namely Graph-Based Visual Saliency (GBVS) \cite{harel2007graph}, Boolean Map Saliency (BMS) \cite{zhang2013saliency}, Unsupervised Hierarchical Model (UHM) \cite{tavakoli2016bottom}, Contextual Encoder-Decoder (CED) \cite{kroner2019contextual} and Saliency GAN (SalGAN) \cite{pan2017salgan}, which are listed as one of the top performing models in the MIT300 Benchmark \cite{bylinskii2015saliency}. The output of the 2D saliency model are local saliency maps, and the different rotations covering the surroundings of the central position should provide the global context of the scene.
\subsection{Input Fusion CNN}
\label{Input Fusion CNN}
The output of the previous step are saliency maps for all the CMP faces. These CMP saliency maps are organized in order to feed the CNN that merges them together to obtain one saliency map by face. This operation is performed by arranging the corresponding faces with similar indexes channel-wise, \textit{i.e.}, the ones that face the same direction but with different rotation. This will provide tensors with a shape of $l \times l \times 5$. 

The fusion method is used to combine the information from different contexts, and to improve the performance of the saliency prediction, especially when the used projection is with a limited field of view. The proposed fusion method is a CNN-based approach. There are several ways to merge information from different sources: input fusion, early fusion, late fusion, ad-hoc fusion and hyper-dense fusion \cite{zhang2019hyperfusion}. We adopted in this work the input fusion CNN approach. This kind of CNN performs fusion channel-wise, where the images are stacked together with the respect of the third dimension before feeding them into the network, and the output is shaped according to one of the stacked images. The training procedure highlighted by the blue dotted lines in the Figure \ref{fig:framework} as well as the architecture of the input fusion CNN block are provided below.
\subsubsection{Architecture}
as shown in Table \ref{tab:cnnarch}, the CNN has six padded convolution layers, with batch normalization after each convolution. The input tensor is with a shape of $l \times l \times 5$, and the output is a single flattened saliency map corresponding to the central CMP rotation (0$^\circ$, 0$^\circ$). 
		\begin{table}[t!]\scriptsize
		\centering
		\caption{The architecture of the input fusion CNN.}
        \begin{tabular}{|c|c|c|c|c|}
        \hline
        \textbf{Name} & \textbf{Kernel size} & \textbf{Input shape} & \textbf{Output shape} & \textbf{Activation} \\ \hline
        Conv1         & (7, 7, 32)           & (512, 512, 5)        & (512, 512, 32)        & ReLU                \\ \hline
        Conv2         & (7, 7, 64)           & (512, 512, 32)       & (512, 512, 64)        & ReLU                \\ \hline
        Conv3         & (7, 7, 128)          & (512, 512, 64)       & (512, 512, 128)       & ReLU                \\ \hline
        Conv4         & (7, 7, 64)           & (512, 512, 128)      & (512, 512, 64)        & ReLU                \\ \hline
        Conv5         & (7, 7, 32)           & (512, 512, 64)       & (512, 512, 32)        & ReLU                \\ \hline
        Conv6         & (7, 7, 1)            & (512, 512, 32)       & (512, 512, 1)         & Sigmoid             \\ \hline
        Flatten       & /                    & (512, 512, 1)        & (262144)              & /                   \\ \hline
        \end{tabular}
        \label{tab:cnnarch}
        \end{table}
\subsubsection{Training}
the CNN has been trained using datasets from the ICME Grand Challanges 2017 and 2018 \cite{rai2017dataset, gutierrez2018introducing}. The details about the training set are provided in Section \ref{sec:experimental}. The loss function $\mathcal{L}$ used in the training process, given by Eq. (\ref{loss}), consists of a linear combination of different saliency evaluation metrics, namely Kullback-Leiber Divergence (KLD), Correlation Coefficient (CC) and Binary Cross Entropy (BCE). As mentionned in \cite{bruckert2019deep}, this combination can lead to significant improvements in performance.
		\begin{equation}
		\mathcal{L}(y, \hat{y}) = \alpha \, KLD(y, \hat{y}) + \beta \, (1-CC(y, \hat{y}))+ \gamma \, BCE(y, \hat{y}),
		\label{loss}
		\end{equation}
where $y$ is the ground truth saliency map that is converted into CMP according to the central rotation  (0$^\circ$, 0$^\circ$), and $\hat{y}$ is the output of the CNN-base fusion block, while $\alpha$, $\beta$ and $\gamma$ are three scalars used to balance the considered saliency evaluation metrics.
\subsection{Post-Processing} 
After obtaining the saliency maps from the Input Fusion CNN block in the CMP format, a series of operations are carried out to project back the faces to the ERP in order to obtain the final saliency map. It has been shown statistically that the viewer tends to direct his attention towards the equator when exploring an omnidirectional scene \cite{de2017look}. Therefore, the introduction of the equator bias is an essential operation to increase the performance. We explored several ways to apply equator bias, and we have obtained the best performance by averaging the ground-truth training data as an equator bias \cite{suzuki2018saliency}. Thus, the estimated saliency map after conversion to ERP format, denoted as $S_{ini}$, is multiplied by the equator bias $EB$, resulting in the final saliency map $S_{Final}$ defined as follows:
	\begin{align}
	\label{eb}
	    S_{Final} &= S_{ini} \,  EB, \nonumber\\
	\text{with}	\quad	EB &= \frac{1}{N}\sum\limits_{k=1}^N GT^{(k)}, 
 	\end{align}
where $GT^{(k)}$ is the $k^{th}$ ground truth saliency map and $N$ is the number of images in the training set.
\section{Experimental Results}
\label{sec:experimental}
To evaluate the performance of our proposed framework, we used the datasets of Salient!360 ICME 2017 \& 2018 Grand Challenges \cite{rai2017dataset, gutierrez2018introducing} for head only prediction, where the salient!360 2017 dataset contains 20 ERP images, and the salient!360 2018 dataset has 85 ERP training images. We converted each ERP image into 6 fixed-size $512\times512$ CMP faces with 5 rotations and stack them channel-wise as described in subsection \ref{Input Fusion CNN}. Thus, we generated $105\times6\times5 = 3150$ training CMP faces. We trained the input fusion CNN model using random initialization, weight regularization, and Adam Optimization algorithm \cite{kingma2014adam} for 20 epochs. Loss function parameters $\alpha$, $\beta$ and $\gamma$ were empirically set to 0.5, 0.25 and 0.25, respectively.

For testing stage, we used the test set of Salient!360 2017 challenge, which consists of 25 ERP images and their corresponding head saliency maps. In order to show the effectiveness of the proposed CNN-based fusion approach, we compared it to the weighted average merge method. For the latter, the same steps as those adopted in the proposed method are used, except that the CMP saliency maps are linearly combined, \textit{i.e.}, average. For this evaluation, we have considered 5 models, namely GBVS \cite{harel2007graph}, BMS \cite{zhang2013saliency}, UHM \cite{tavakoli2016bottom}, CED \cite{kroner2019contextual} and SalGAN \cite{pan2017salgan}. Table \ref{results} shows clearly that the simple linear combination of the CMP saliency maps, as is widely performed in the literature, does not achieve optimal results. In contrast, the proposed method clearly outperforms the weighted average merge method, for all the considered 2D saliency models. The obtained results demonstrate that the increase in performance is not mainly related to the used 2D saliency models, but thanks to the use of CNN-based fusion of stacked CMP saliency maps.

Given that the best performance have been achieved with SalGAN method. We compared these results against the state-of-the-art head movement prediction models. Specifically, we considered the top-\textit{5} methods \cite{startsev2018360,lebreton2018gbvs360,zhu2018prediction,ling2018saliency,battisti2018feature} from Salient!360 2017 and we used the same evaluation metrics $KLD$ and $CC$. The reported result in table \ref{results2} shows the high efficiency of our framework compared to the stat-of-the-art methods.
\begin{table}[t!]\scriptsize
\centering
\caption{Performance evaluation of the saliency maps merge methods.}
\begin{tabular}{c|l|c|c|}
\hline
\multicolumn{1}{|l|}{\textbf{Saliency Model}} & \multicolumn{1}{c|}{\textbf{Fusion Method}} & \textbf{KLD$\downarrow$}        & \textbf{CC$\uparrow$} \\ \hline
\multicolumn{1}{|c|}{\multirow{2}{*}{\textbf{CED}}}    & \multicolumn{1}{c|}{Input Fusion CNN} & 0.71                & 0.74                       \\ \cline{2-4} 
\multicolumn{1}{|c|}{}                                 &\multicolumn{1}{c|}{Average method}                               & 1.01                & 0.72    \\ \hline\hline
\multicolumn{1}{|c|}{\multirow{2}{*}{\textbf{UHM}}}    & Input Fusion CNN                      & 0.47                & 0.74                     \\ \cline{2-4} 
\multicolumn{1}{|c|}{}                                 & \multicolumn{1}{c|}{Average method}          & 0.51                & 0.72                      \\ \hline\hline
\multicolumn{1}{|c|}{\multirow{2}{*}{\textbf{BMS}}}    & Input Fusion CNN                      & 0.99                & 0.66                      \\ \cline{2-4} 
\multicolumn{1}{|c|}{}                                 & \multicolumn{1}{c|}{Average method}                               & 2.57                & 0.48                       \\ \hline\hline
\multicolumn{1}{|c|}{\multirow{2}{*}{\textbf{GBVS}}}   & \multicolumn{1}{c|}{Input Fusion CNN} & 0.49                & 0.75                          \\ \cline{2-4} 
\multicolumn{1}{|c|}{}                                 &\multicolumn{1}{c|}{Average method}                              & 0.46                    & 0.73                        \\ \hline\hline
\multicolumn{1}{|c|}{\multirow{2}{*}{\textbf{SalGAN}}} & Input Fusion CNN                      & 0.47               & 0.76                           \\ \cline{2-4} 
\multicolumn{1}{|c|}{}                                 & \multicolumn{1}{c|}{Average method}          & 1.84               & 0.64                          \\ \hline
\end{tabular}
\label{results}
\end{table}

\begin{table}[t!]
\centering
\caption{Comparison against existing head movement prediction methods. The two best results are in bold.}
\begin{tabular}{|l|c|c|}
\hline
\textbf{Method} & \textbf{KLD$\downarrow$}        & \textbf{CC$\uparrow$}       \\ \hline
Startsev \textit{et al.} \cite{startsev2018360}    & 0.75                & 0.62                  \\ \hline
Lebreton \textit{et al.} \cite{lebreton2018gbvs360}   & \textbf{0.44}                & 0.69                \\ \hline
Zhu \textit{et al.} \cite{zhu2018prediction}  & 0.65                & 0.67              \\ \hline

Ling \textit{et al.} \cite{ling2018saliency}  & 0.51                & \textbf{0.71}              \\ \hline
Battisti \textit{et al.} \cite{battisti2018feature}  & 0.81                & 0.52              \\ \hline
Our method    & \textbf{0.47}                & \textbf{0.76}           \\ \hline
\end{tabular}
\label{results2}
\end{table}
\section{Conclusion}
\label{sec:conclusion}
This paper presented a novel framework to extend any 2D saliency prediction method for 360-degree images. Our contribution mainly focused on the fusion method of CMP saliency maps using CNN-based approach. The proposed method has been compared to the weighted average merge approach using five 2D saliency methods. The results clearly showed the effectiveness and the added value of the proposed fusion block.

In addition, the modularity of the proposed framework allows for improving it for better performance, especially if it is combined with high performing 2D saliency methods and other enhancements in the rest of the chain. These points will be the subject of future contributions.
\bibliographystyle{IEEEbib}
\bibliography{strings}
\end{document}